\documentclass[12pt,a4paper]{article}
\usepackage{t1enc}
\usepackage[latin1]{inputenc}
\usepackage[english]{babel}
\usepackage{amssymb,amsmath}
\usepackage{graphicx}
\usepackage{hyperref}
\usepackage{epstopdf}

\begin{document}

\title{ Three errors in the article:" The OPERA neutrino velocity result and the synchronisation of clocks, arXiv:1109.6160"}
\author{Olivier Besida \\  CEA, Irfu, SPP \\  Centre de Saclay, F-91191 Gif-sur-Yvette, France
\\olivier.besida@cea.fr}
\maketitle

\begin{abstract} 
We found three mistakes in the article " The OPERA neutrino velocity result and the synchronisation of clocks" by Contaldi \cite{Contaldi}. First, the definition of the angle of the latitude in the geoid description leads to a prolate spheroid (rugby ball shape) instead of an oblate spheroid with the usual equatorial flattening. Second, Contaldi forgot a cosine of the latitude in the centripetal contribution term. And last but not least, a profound conceptual mistake was done in believing that an atomic clock or any timekeeper apparatus was carried in a journey by car or plane between CERN and Gran Sasso; instead of that atomic clocks are continuously resynchronized through a GPS device, and the variation of the potential term applies only for the neutrino travel itself. Thus instead of a $\Delta t \approx 30ns $ correction claimed by the author in a travel of 12 hours plus 4 days at rest for an atomic clock, we have found a time correction only for the neutrino itself  $\Delta t=3.88 \, 10^{-16} s$! That means, that this paper \cite{Contaldi} does not give the right explanation why the neutrino is seen travelling faster than the speed of light in the OPERA neutrino experiment. 
\end{abstract}

\section{Introduction}
Last paper of the OPERA collaboration \cite{OPERA} generates a lot of reactions and comments about the observation of the advance of 60.7 ns of the neutrino beam coming from the CERN compared to the time of travel given by the speed of light. One of these comments was given by Contaldi \cite{Contaldi}, explaining the advance by an effect of general relativity for travelling atomic clocks. Three errors and misunderstanding by the author of this article \cite{Contaldi} are detailed here: one is an angular definition, the second is a missing term and third a misunderstanding of the method used by OPERA and CERN to continuously resynchronize their atomic clocks with GPS devices. Thus the supposed correction of $\Delta t \approx 30 \,ns $  \cite{Contaldi} is shown to be irrelevant.

\section{Error 1 : On geoid description}
C. R. Contaldi in his paper arXiv:1109.6160 \cite{Contaldi} uses a polynomial expansion
up to the quadrupolar term of the geoid model.
This type of geoid model is quite usual except the fact an error 
was introduced in the angular definition. Contaldi states :

 \begin{align}
&\ \theta = latitude\notag \\
&\ x=sin(\theta)  \notag \\
&\ r \approx 6356742.025 +21353.642x^2+39.832x^4\notag \\
\end{align}

\begin{figure}[htp] %on ouvre l'environnement figure
\centering
\includegraphics[width=14cm,height=15cm]{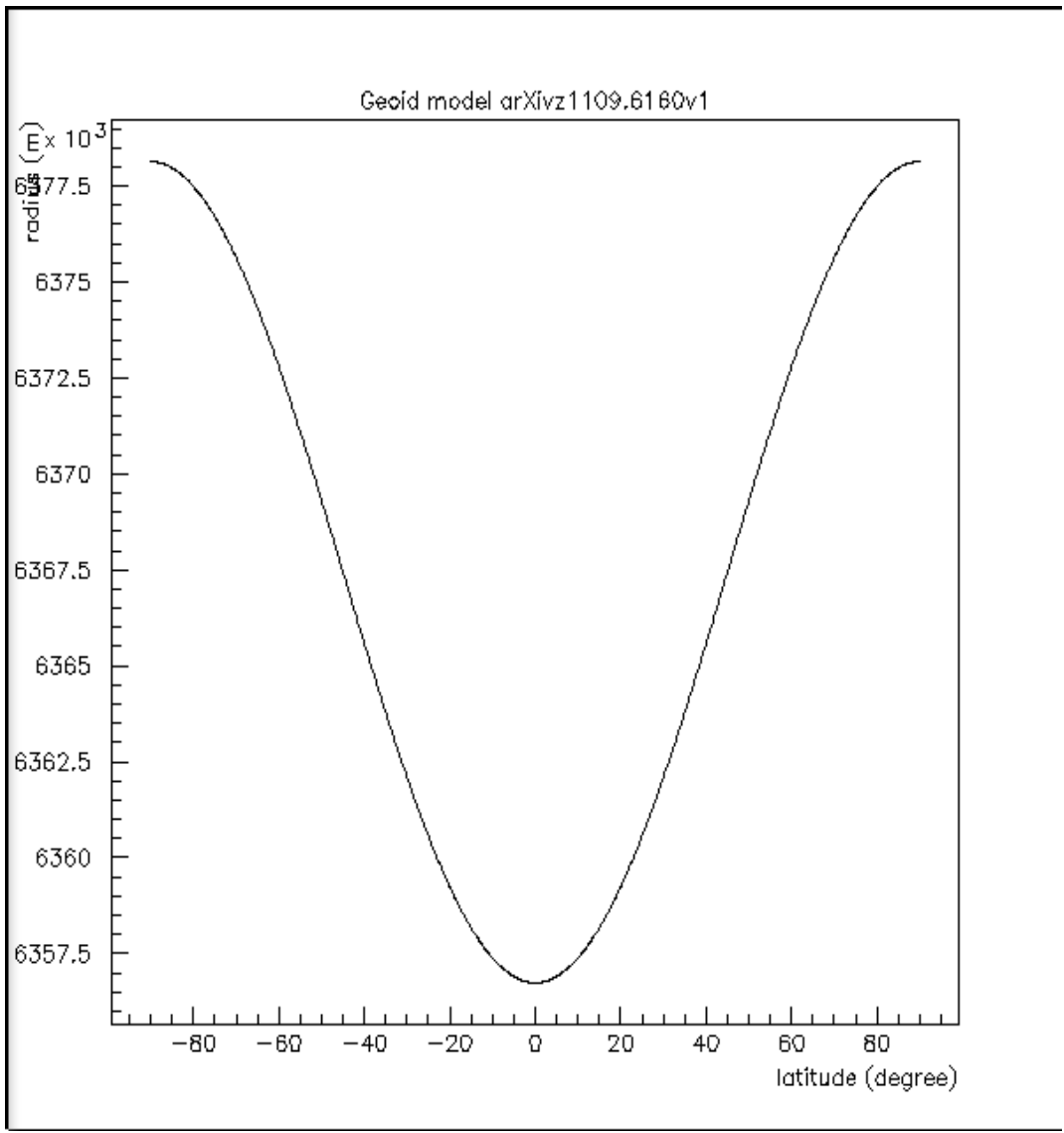}
\caption{Contaldi's prolate geoid} %la légende
\label{geoidContaldi} %l'étiquette pour faire référence à cette image
\end{figure}

Such a model describes a prolate spheroid (rugby balloon shape), with an equatorial radius smaller than the polar radius (see fig \ref{geoidContaldi} ). Unfortunately, we knew for few centuries that the surface of the Earth does not have this shape (cf Newton 1687, Huyghens 1690, La Condamine, Bouguer and Godin 1736, Clairaut 1736, Maupertuis 1742,  Maclaurin 1742).
 The equatorial radius of the Earth is greater than the radii at the poles. The famous equatorial flatenning of the Earth gives at this order of polynomial expansion an oblate shape to the Earth. In WGS-84 referential, the adopted value of the flattening of the Earth is 1:298.257223563.

Thus a correct description of the earth would be \cite{Ashby}:

 \begin{align}
&\ \lambda = latitude \notag \\
&\ \theta = {\pi \over 2}-\lambda \notag \\
&\ x=sin(\theta)   \notag \\
&\ r\approx 6356742.025 +21353.642x^2+39.832x^4 \notag \\
\end{align}

\section{ Error 2 : Correction of the centripetal potential term}

An effective gravitational potential of the Earth including a centripetal contribution and a quadrupolar contribution (see eq. 2) \cite{Contaldi} is described .
Contaldi states:

 \begin{align}
&\ V = - {{G M_E}\over r}[1-{J_2 ({r_E \over r } )^2 P_2(cos(\theta)}]+{{1 \over 2}(\omega_E r)^2} 
\end{align}

A mistake was done in the second term of this expression, as the radius $ r $ used here is not the distance  to the center of the Earth but to the axis of rotation of the Earth. The centripetal correction to the effective potential should be null at the poles and is maximal at the equator of the Earth. This can be corrected inserting a cosine of the latitude (or $sin(\theta)$) in the second term of the potential as follows:
 \begin{align}
&\ \theta = {\pi \over 2}-latitude \notag \\
&\ V = - {{G M_E}\over r}[1-{J_2 ({r_E \over r } )^2 P_2(cos(\theta)}]+{{1 \over 2}(\omega_E r sin(\theta))^2} 
\end{align}

Thus, assuming the following latitude for the CERN : $ 46.235 ° N $ and for the Gran Sasso Laboratory (LNGS) : $ 42.420 ° N $ , we find :
 \begin{align}
&\ r_{CERN}=6366967.8 \, m \notag \\
&\ r_{LNGS}=6256742.0 \, m \notag \\
&\ {V_{CERN} \over c^{2}}=-6.95787\, 10^{-10} \notag \\
&\ {V_{LNGS} \over c^{2}} = -6.95627\, 10^{-10} \notag \\
&\ {\Delta V \over c^{2}} = 1.60 \, 10^{-13} \notag \\
\end{align}

So while the author of \cite{Contaldi} found a variation of the effective potential correction term between CERN and LNGS $ {\Delta V \over c^2} = 7.82 \, 10^{-14} $ , taking into account theses two errors we have computed here a value: $ {\Delta V \over c^{2}} = 1.60 \, 10^{-13} $ , which is twice the value previously found by Contaldi \cite{Contaldi}.

\section{Error 3 : a misunderstanding, no atomic clock was transported from CERN to Gran-Sasso for synchronization, except neutrinos}

 And last but not least, a profound conceptual mistake was done believing that an atomic clock or any timekeeper apparatus was carried in a journey by car or plane between CERN to Gran Sasso as it was switched on and keeping the time during the travel. As explained by the OPERA Collaboration  \cite{OPERA} the atomic clocks used either at CERN or at Gran Sasso are continuously resynchronized through two GPS devices at each location with the satellites. Moreover these two GPS devices were independently calibrated by METAS and PTB for common view mode time transfert \cite{OPERA} .
 Consequently, the variation of potential term applies only for the time correction of the travel of the neutrino itself. As the CERN-LNGS distance is about 730 km, this time corrections applies to 2.43 millisecond instead of 12 hours + 4 days at rest as mentioned in Contaldi's scenario. 
 Thus instead of a $\Delta t \approx 30  \, ns $ correction claimed by Contaldi in a travel of 12 hours plus 4 days at rest for an atomic clock, we have found a time correction to be applied only on the neutrino travel itself  $\Delta t=3.88 \, 10^{-16} s$.
 Of course, such a time correction is far away from the sensitivity of the OPERA experiment, and it cannot explain any advance of the detected neutrinos of 60.7 ns as observed by OPERA \cite{OPERA}.

\section{Conclusion}
Three important errors are found in the paper of C.R. Contaldi \cite{Contaldi} trying to explain some time corrections in the synchronization of clocks for the neutrino beam travelling from CERN to LNGS as observed by the OPERA experiment. One first error is due to a wrong definition of a polar angle related to the latitude in the calculation of the geoid model of the surface of the earth, leading to rugby ball shape instead of an oblate shape with an equatorial flattening. A second error in the paper arXiV:1109.6160, is due to the missing of a sine term of the polar angle in the centripetal correction term. And third, a conceptual error appeared in believing that an atomic clock has been carried at the surface of the Earth by car or plane during several hours or days between CERN and LNGS in order to synchronize the atomic clocks. Instead of that, the atomic clocks are continuously synchronized thanks to GPS satellites at a high level of precision( see  \cite{OPERA} ). Thus the time correction of $\Delta t \approx 30 \, ns $  claimed by Contaldi,  is reduced to  $\Delta t=3.88 \, 10^{-16} s$.
So the mystery of the neutrino faster the speed of light remains complete.

\section{Acknowledgement}
I must thanks J. Rich, B. Vallage and G. Vasseur for fruitful discussions about the measurement of the speed of the neutrino from CERN observed at LNGS by the OPERA collaboration.

\begin{thebibliography}{3}

\bibitem{Contaldi}  C.R. Contaldi, {\em The OPERA neutrino velocity result and the synchronisation of clocks}, arXiV:1109.6160

\bibitem{OPERA} The OPERA Collaboration: T. Adam et al., {\em Measurement of the neutrino velocity with the OPERA detector in the CNGS beam ,arXiv:1109.4897}

\bibitem{Ashby} Neil Ashby, {\em Relativity in the Global Positioning System, p.13}
\url{http://physics.colorado.edu/faculty/ashby_n.html}

\end{thebibliography}
\end{document}